\documentclass[twocolumn,showpacs,amsmath,amssymb,superscriptaddress]{revtex4}
\usepackage{graphicx}
\usepackage{bm}
\newcommand{\gsim}{\;\rlap{\lower 3.5 pt \hbox{$\mathchar \sim$}} \raise 1pt \hbox {$>$}\;}
\newcommand{\lsim}{\;\rlap{\lower 3.5 pt \hbox{$\mathchar \sim$}} \raise 1pt \hbox {$<$}\;}
\begin{document}
\title{\boldmath Updated NNLO QCD Predictions for the Weak Radiative $B$-Meson Decays}

\author{M.~Misiak}
\affiliation{Institute of Theoretical Physics, University of Warsaw, PL-02-093 Warsaw, Poland}
\author{H.~M.~Asatrian}
\affiliation{Yerevan Physics Institute, 0036 Yerevan, Armenia}
\author{R.~Boughezal}
\affiliation{High Energy Physics Division, Argonne National Laboratory, Argonne, Illinois 60439, USA}
\author{M.~Czakon}
\affiliation{Institut f\"ur Theoretische Teilchenphysik und Kosmologie, RWTH Aachen University, D-52056 Aachen, Germany}
\author{T.~Ewerth}
\affiliation{Institut f\"ur Theoretische Teilchenphysik, Karlsruhe Institute of Technology, D-76128 Karlsruhe, Germany}
\author{A.~Ferroglia}
\affiliation{New York City College of Technology, CUNY, Brooklyn, New York 11201, USA}
\affiliation{The Graduate School and University Center, CUNY, New York, New York 10016, USA}
\author{P.~Fiedler}
\affiliation{Institut f\"ur Theoretische Teilchenphysik und Kosmologie, RWTH Aachen University, D-52056 Aachen, Germany}
\author{P.~Gambino}
\affiliation{Dipartimento di Fisica, Universit\`a di Torino \&  INFN, Torino, I-10125 Torino, Italy}
\author{C.~Greub}
\affiliation{Albert Einstein Center for Fundamental Physics, Institute for Theoretical Physics, University of Bern, CH-3012 Bern, Switzerland}
\author{U.~Haisch}
\affiliation{Rudolf Peierls Centre for Theoretical Physics,
    University of Oxford, OX1 3PN Oxford, United Kingdom}
\affiliation{CERN, Theory Division, CH-1211 Geneva 23, Switzerland}
\author{T.~Huber}
\affiliation{Theoretische Physik 1, Naturwissenschaftlich-Technische Fakult\"at, Universit\"at Siegen, D-57068 Siegen, Germany}
\author{M.~Kami\'nski}
\affiliation{Institute of Theoretical Physics, University of Warsaw, PL-02-093 Warsaw, Poland}
\author{G.~Ossola}
\affiliation{New York City College of Technology, CUNY, Brooklyn, New York 11201, USA}
\affiliation{The Graduate School and University Center, CUNY, New York, New York 10016, USA}
\author{M.~Poradzi\'nski}
\affiliation{Institute of Theoretical Physics, University of Warsaw, PL-02-093 Warsaw, Poland}
\affiliation{Theoretische Physik 1, Naturwissenschaftlich-Technische Fakult\"at, Universit\"at Siegen, D-57068 Siegen, Germany}
\author{A.~Rehman}
\affiliation{Institute of Theoretical Physics, University of Warsaw, PL-02-093 Warsaw, Poland}
\author{T.~Schutzmeier}
\affiliation{Physics Department, Florida State University, Tallahassee, Florida 32306-4350, USA}
\author{M.~Steinhauser}
\affiliation{Institut f\"ur Theoretische Teilchenphysik, Karlsruhe Institute of Technology, D-76128 Karlsruhe, Germany}
\author{J.~Virto}
\affiliation{Theoretische Physik 1, Naturwissenschaftlich-Technische Fakult\"at, Universit\"at Siegen, D-57068 Siegen, Germany}

\begin{abstract}

Weak radiative decays of the $B$ mesons belong to the most important
flavor changing processes that provide constraints on physics at the TeV
scale.  In the derivation of such constraints, accurate standard model
predictions for the inclusive branching ratios play a crucial role.  In the
current Letter we present an update of these predictions, incorporating
all our results for the ${\mathcal O}(\alpha_s^2)$ and lower-order
perturbative corrections that have been calculated after 2006. New estimates
of nonperturbative effects are taken into account, too.  For the $CP$- and
isospin-averaged branching ratios, we find
${\mathcal B}_{s\gamma} = (3.36 \pm 0.23) \times 10^{-4}$
and
${\mathcal B}_{d\gamma} = \left(1.73^{+0.12}_{-0.22}\right) \times 10^{-5}$,
for $E_\gamma > 1.6\,$GeV. Both results remain in agreement with the current
experimental averages. Normalizing their sum to the inclusive semileptonic
branching ratio, we obtain
$R_\gamma \equiv 
\left({\mathcal B}_{s\gamma}+{\mathcal B}_{d\gamma}\right)/{\mathcal B}_{c\ell\nu}
= (3.31 \pm 0.22) \times 10^{-3}$. 
A new bound from ${\mathcal B}_{s\gamma}$ on the charged Higgs boson mass in
the two-Higgs-doublet model II reads 
  $M_{H^{\pm}} > 480\,$GeV at 95\%C.L. 

\end{abstract}

                             
\maketitle

\section{Introduction \label{sec:intro}}

The inclusive decays $\bar B \to X_s\gamma$ and $\bar B \to X_d\gamma$ are
considered among the most interesting flavor changing neutral current
processes.  They contribute in a significant manner to current bounds on
masses and interactions of possible additional Higgs bosons and/or
supersymmetric particles. The evaluation of such bounds depends in a crucial
manner on both the central values and uncertainties of the branching ratio
predictions within the standard model (SM). Updating the SM predictions is the
main purpose of the present Letter.

Measurements of the $CP$- and isospin-averaged $\bar B \to X_s\gamma$ branching
ratio by CLEO~\cite{Chen:2001fj}, Belle~\cite{Abe:2001hk,Limosani:2009qg}, and
BABAR~\cite{Lees:2012ym,Lees:2012ufa,Lees:2012wg,Aubert:2007my} lead to the
combined result~\cite{Amhis:2014hma}
\begin{equation}
{\mathcal B}^{\rm exp}_{s\gamma} = (3.43 \pm 0.21 \pm 0.07)\times 10^{-4},
\end{equation}
for the photon energy $E_\gamma > E_0 = 1.6\,$GeV in the decaying meson
rest frame. The combination involves an extrapolation from measurements
performed at $E_0 \in [1.7,2.0]\,$GeV. Applying the same extrapolation method
to the available $\bar B \to X_d\gamma$
measurement~\cite{delAmoSanchez:2010ae}, one finds
\begin{equation}
{\mathcal B}^{\rm exp}_{d\gamma} = (1.41 \pm 0.57)\times 10^{-5}
\end{equation}
at $E_0 = 1.6\,$GeV~\cite{Crivellin:2011ba}. More precise determinations of
${\mathcal B}^{\rm exp}_{q\gamma}$ for $q=s,d$ are expected from
Belle~II~\cite{Aushev:2010bq}.

Theoretical calculations of ${\mathcal B}_{q\gamma}$ have a chance to match
the experimental precision only in a certain range of $E_0$ where the
nonperturbative contribution $\delta \Gamma_{\rm nonp}$ in the relation 
\begin{equation}
\Gamma(\bar B \to X_q \gamma) ~=~ \Gamma(b \to X_q^p \gamma) ~+~ \delta \Gamma_{\rm nonp}
\end{equation}
remains under control. Here, $\Gamma(b \to X_q^p \gamma)$ denotes the
perturbatively calculable rate of the radiative $b$-quark decay involving only
charmless partons in the final state. Their overall strangeness vanishes for
$X_d^p$ and equals $-1$ for $X_s^p$. The analysis of
Ref.~\cite{Benzke:2010js} implies that unknown contributions to $\delta
\Gamma_{\rm nonp}$ are potentially larger than the so-far determined ones, and
induce around $\pm5\%$ uncertainty in ${\mathcal B}_{s\gamma}$ at $E_0 =
1.6\,$GeV. Nonperturbative uncertainties in ${\mathcal B}_{d\gamma}$ receive
additional sizeable contributions~\cite{Asatrian:2013raa} due to collinear
photon emission in the $b \to du\bar u\gamma$ process whose
Cabibbo-Kobayashi-Maskawa (CKM) factor is only a few times smaller than
the one in the leading term.

Apart from possible future progress in analyzing nonperturbative effects, one
needs to determine $\Gamma(b \to X_q^p \gamma)$ to a few percent accuracy. It
requires evaluating next-to-next-to-leading order (NNLO) QCD
corrections that involve Feynman diagrams up to four loops. The first 
SM estimate of the $\bar B \to X_s \gamma$ branching
ratio at this level was presented in Ref.~\cite{Misiak:2006zs} where all the
corrections calculated up to 2006 were taken into account. A part of the
${\mathcal O}(\alpha_s^2)$ contribution was obtained via
interpolation~\cite{Misiak:2006ab} in the charm quark mass between the
large-$m_c$ asymptotic expression~\cite{Misiak:2010sk} and the $m_c=0$
boundary condition that was estimated using the Brodsky-Lepage-Mackenzie (BLM)
approximation~\cite{Brodsky:1982gc}.

In the present Letter, we provide an updated prediction for ${\mathcal
B}_{s\gamma}$, including all the contributions and estimates worked out after
the completion of Ref.~\cite{Misiak:2006zs}. They are listed in
Sec.~\ref{sec:bsg} where the necessary definitions are introduced. The
interpolation in $m_c$ is still being applied. However, the $m_c=0$ boundary
condition is no longer a BLM-based estimate but rather comes from an explicit
calculation~\cite{Czakon:2015exa}.

The current analysis supersedes our previous one in
Ref.~\cite{Misiak:2006zs}, which was published in 2006 and has not been updated
since then. It has been widely considered as a standard reference until
now. The time for our update comes only at present because the most recent and
technically challenging four-loop calculation of Ref.~\cite{Czakon:2015exa}
constitutes a breakthrough in the analysis. It has an important effect on the
central value of ${\mathcal B}_{s\gamma}$.

The Letter is organized as follows. After discussing ${\mathcal B}_{s\gamma}$
in Sec.~\ref{sec:bsg}, our NNLO analysis is extended to ${\mathcal
B}_{d\gamma}$ in Sec.~\ref{sec:bdg}. Next, in Sec.~\ref{sec:rg}, we consider
$R_\gamma \equiv \left({\mathcal B}_{s\gamma}+{\mathcal B}_{d\gamma}\right)/{\mathcal B}_{c\ell\nu}$
which may sometimes be more convenient than ${\mathcal B}_{s\gamma}$ for
deriving constraints on new physics.  Sec.~\ref{sec:np} is devoted to
presenting a generic expression for beyond-SM contributions, as well as an
updated bound for the charged Higgs boson mass in the 
two-Higgs-doublet model II (THDM~II). We conclude
in Sec.~\ref{sec:sum}.

\section{\boldmath ${\mathcal B}_{s\gamma}$ in the SM \label{sec:bsg}}

Radiative $B$-meson decays are most conveniently described in the framework
of an effective theory that arises after decoupling of the $W$ boson and
heavier particles. Flavor-changing weak interactions that are relevant for
$\Gamma(b \to X_q^p \gamma)$ with $q=s,d$ are given by 
\begin{equation} \label{Leff}
{\mathcal L}_{\rm eff} ~\sim~ V_{tq}^* V_{tb} \left[ \sum_{i=1}^8 C_i Q_i 
+ \kappa_q \sum_{i=1}^2 C_i (Q_i-Q_i^u)\right].
\end{equation}
Explicit expressions for the current-current ($Q_{1,2}$), four-quark penguin
($Q_{3,\ldots,6}$), photonic dipole ($Q_7$), and gluonic dipole ($Q_8$)
operators can be found, e.g., in Eq.~(2.5) of Ref.~\cite{Misiak:2006ab}. The
CKM element ratio $\kappa_q = (V_{uq}^* V_{ub})/(V_{tq}^* V_{tb})$ is small
for $q=s$, and it affects ${\mathcal B}_{s\gamma}$ by less than $0.3\%$. Barring
this effect and the higher-order electroweak ones, $\Gamma(b \to X_s^p
\gamma)$ in the SM is given by a quadratic polynomial in the real Wilson
coefficients $C_i$
\begin{equation}
\Gamma(b \to X_s^p \gamma) ~\sim~ \sum_{i,j=1}^8 C_i C_j\; G_{ij}.
\end{equation}
A series of contributions to the above expression from our calculations in
Refs.~\cite{Czakon:2015exa,Czakon:2006ss, Asatrian:2006rq, Ewerth:2008nv,
Boughezal:2007ny, Asatrian:2010rq, Ferroglia:2010xe,Misiak:2010tk,
Kaminski:2012eb, Huber:2014nna} makes the current analysis significantly
improved with respect to the one in Ref.~\cite{Misiak:2006zs}. In particular,
the NNLO Wilson coefficient calculation becomes complete after including the
four-loop anomalous dimensions that describe $Q_{1,\ldots,6} \to Q_8$ mixing
under renormalization~\cite{Czakon:2006ss}. Effects of the charm and bottom
quark masses in loops on the gluon lines in $G_{77}$~\cite{Asatrian:2006rq},
$G_{78}$~\cite{Ewerth:2008nv} and $G_{(1,2)7}$~\cite{Boughezal:2007ny}, as
well as a complete calculation of $G_{78}$~\cite{Asatrian:2010rq}, are now
available. Three- and four-body final-state contributions to
$G_{88}$~\cite{Ferroglia:2010xe,Misiak:2010tk} and
$G_{(1,2)8}$~\cite{Misiak:2010tk} are included in the BLM approximation.
Four-body final-state contributions involving the penguin and $Q^u_{1,2}$
operators are taken into account at the leading order
(LO)~\cite{Kaminski:2012eb} and next-to-leading order
(NLO)~\cite{Huber:2014nna}. Last but not least, the complete NNLO
calculation~\cite{Czakon:2015exa} of $G_{17}$ and $G_{27}$ at $m_c=0$ is used
as a boundary for interpolating their unknown parts in $m_c$.

Following the algorithm described in detail in Ref.~\cite{Czakon:2015exa},
taking into account new nonperturbative
effects~\cite{Benzke:2010js,Ewerth:2009yr,Alberti:2013kxa}, as well as the
previously omitted parts of the NNLO BLM
corrections~\cite{Ligeti:1999ea}, we arrive at the following SM prediction
\begin{equation} \label{bsgsm}
{\mathcal B}^{\rm SM}_{s\gamma} ~=~ (3.36 \pm 0.23) \times 10^{-4}
\hspace{5mm} \mbox{for~}E_0=1.6\,\mbox{GeV}.
\end{equation}
Individual contributions to the total uncertainty are of nonperturbative
($\pm 5\%$), higher-order ($\pm 3\%$), interpolation ($\pm 3\%$) and
parametric ($\pm 2\%$) origin. They are combined in quadrature.  The
parametric one gets reduced with respect to Ref.~\cite{Misiak:2006zs}, which
becomes possible thanks to the new semileptonic fits of
Ref.~\cite{Alberti:2014yda}. Our input parameters, their uncertainties and
the corresponding correlation matrix can be found in Appendix~D of
Ref.~\cite{Czakon:2015exa}. Since we normalize to the semileptonic branching
ratio ${\mathcal B}_{c\ell\nu}$, our result shows little sensitivity to the
$b$-quark mass and the CKM angles. The main parametric uncertainty ($\pm
1.5\%$) originates from ${\mathcal B}_{c\ell\nu}$, while the next one
($\pm 0.75\%$) comes from $\alpha_s(M_Z)$.

As far as the interpolation uncertainty is concerned, one might have
hoped for its reduction with respect to Ref.~\cite{Misiak:2006zs} after the
explicit evaluation of the $m_c=0$ boundary~\cite{Czakon:2015exa}.
Unfortunately, the interpolated parts of the ${\mathcal O}(\alpha_s^2)$
contributions to $G_{(1,2)7}$ turn out to be sizeable. Their effect on
${\mathcal B}^{\rm SM}_{s\gamma}$ grows from 0 to around 5\% when $m_c$
changes from 0 up to the measured value (see Fig.~4 of
Ref.~\cite{Czakon:2015exa}). In such a situation, we prefer to stay
conservative, and retain our interpolation uncertainty estimate at the
$\pm 3\%$ level.

For the higher-order uncertainty estimation, it is useful to study how
${\mathcal B}^{\rm SM}_{s\gamma}$ depends on three renormalization scales: the
matching scale $\mu_0 \sim m_t$ at which the heavy particles ($t$, $W$, $Z$,
$H^0$) are decoupled, the low-energy scale $\mu_b \sim m_b/2$ at which the
Wilson coefficient renormalization group evolution is terminated, and the
scale $\mu_c$ at which the charm quark mass is renormalized. We vary them in
the ranges $\mu_0 \in [80,320]\,$GeV and $\mu_b,\mu_c\in [1.25,5]\,$GeV,
setting the central values to $\mu_0=160\,$GeV and $\mu_b=\mu_c=2\,$GeV. The
observed scale dependence (see Fig.~6 of Ref.~\cite{Czakon:2015exa}) turns out
to be quite similar to the one in Fig.~2 of
Ref.~\cite{Misiak:2006zs}. Therefore, we leave the higher-order uncertainty
estimate at the $\pm 3\%$ level, i.e., unchanged with respect to
Ref.~\cite{Misiak:2006zs}.

The nonperturbative uncertainty estimate of $\pm 5\%$ is adopted from
Ref.~\cite{Benzke:2010js} without any modification. It turns out to be
identical to our earlier rough estimate in Ref.~\cite{Misiak:2006zs}. Some
comments on possible future suppression of this uncertainty are given in
Sec.~\ref{sec:sum}.

The central value in Eq.~(\ref{bsgsm}) is considerably higher than $3.15
\times 10^{-4}$ in Ref.~\cite{Misiak:2006zs}, although the difference between
the two values does not exceed the previously estimated uncertainty. A
detailed description of various contributions to this difference is given in Sec.~4 
of Ref.~\cite{Czakon:2015exa}, as well as in Table~2 there.

\section{\boldmath ${\mathcal B}_{d\gamma}$ in the SM \label{sec:bdg}} 

Extending our NNLO calculation to the ${\mathcal B}_{d\gamma}$ case begins
with inserting the proper CKM factors in Eq.~(\ref{Leff}). Contrary to
$\kappa_s$, the ratio $\kappa_d$ is not numerically small. Using the CKM fits
of Ref.~\cite{Charles:2015gya}, one finds
\begin{equation}
\kappa_d ~=~ \left(0.007^{+0.015}_{-0.011}\right) + i \left(-0.404^{+0.012}_{-0.014}\right).
\end{equation}
The small real part implies that the effects of $\kappa_d$ on the $CP$-averaged
${\mathcal B}_{d\gamma}$ are dominated by those proportional to
$|\kappa_d|^2$. In such terms, perturbative two- and three-body
final state contributions arise only at the NNLO and NLO, respectively.
They vanish in the $m_c=m_u$ limit, which effectively makes them
suppressed by $m_c^2/m_b^2 \lsim 0.1$. In consequence, the main $\kappa_d$-effect
comes from $b \to du\bar u\gamma$ at the LO, where phase-space suppression
is partially compensated by the collinear logarithms. 

In the first (rough) approximation, one evaluates the tree-level $b \to du\bar
u\gamma$ diagrams retaining a common light-quark mass $m_q$ inside the
collinear logarithms~\cite{Misiak:2010tk}, and varying $m_b/m_q$ between $10
\sim m_B/m_K$ and $50 \sim m_B/m_\pi$ to estimate the uncertainty. The
considered effect varies then from 2\% to 11\% of ${\mathcal B}_{d\gamma}$.  A
more involved analysis with the help of fragmentation functions gives a very
similar range~\cite{Asatrian:2013raa}. Including this contribution
in our evaluation of the entire $B_{d\gamma}$ from Eq.~(\ref{Leff}), we find
\begin{equation} \label{bdgsm}
{\mathcal B}^{\rm SM}_{d\gamma} = \left(1.73^{+0.12}_{-0.22}\right) \times 10^{-5}
\hspace{5mm} \mbox{for~}E_0=1.6\,\mbox{GeV},
\end{equation}
where the central value corresponds to $m_b/m_q=50$. Our result is about 12\%
larger than the one given in Ref.~\cite{Crivellin:2011ba} where
the $b \to du\bar u\gamma$ contributions were neglected. The uncertainty
estimate in Eq.~(\ref{bdgsm}) improves with respect to
Ref.~\cite{Crivellin:2011ba} thanks to including the NNLO QCD corrections and
using the updated CKM fit~\cite{Charles:2015gya}. Interestingly, the
parametric uncertainty due to the CKM input amounts to $\pm 2.5\%$ only.

The collinear logarithm problem might seem artificial because isolated photons
are required in the experimental signal sample. Unfortunately, requiring
photon isolation on the perturbative side would necessitate introducing an
infrared cutoff on the gluon energies, e.g., in the NLO corrections to the 
dominant $G_{77}$ term. Without a dedicated analysis (which is beyond
the scope of the present Letter), it is hard to verify whether such an approach
would enhance or suppress the uncertainty in ${\mathcal B}_{d\gamma}$.

Another question concerning the $|\kappa_d|^2$-terms is whether the off-shell
light vector meson conversion to photons can be assumed to be included in our
overall $\pm 5\%$ nonperturbative uncertainty. Much smaller effects
found in the vector-meson-dominance analysis of Ref.~\cite{Ricciardi:1995jh}
imply that it is likely to be the case.

\section{\boldmath The ratio $R_\gamma$ \label{sec:rg}}

In the fully inclusive measurements of radiative $B$-meson
decays~\cite{Chen:2001fj,Limosani:2009qg,Aubert:2007my,Lees:2012ym,Lees:2012ufa},
the final hadronic state strangeness is not verified. The actually measured
quantity is ${\mathcal B}_{s\gamma}+{\mathcal B}_{d\gamma}$. Next, the result
is divided by
$\left(1+|(V_{td}^* V_{tb})/(V_{ts}^* V_{tb})|^2\right)$ 
to obtain ${\mathcal B}_{s\gamma}$. To avoid such a complication,
we provide here our SM prediction for~ ${\mathcal B}_{s\gamma}+{\mathcal
B}_{d\gamma}$~ with all the correlated uncertainties properly taken into
account. Moreover, we normalize it to the $CP$- and isospin-averaged inclusive
semileptonic branching ratio ${\mathcal B}_{c\ell\nu}$. In the ${\mathcal
B}_{s\gamma}$ case, such a normalization reduces the parametric uncertainty
from $\pm 2.0\%$ to $\{+1.2,-1.4\}\%$. It may also be useful on the experimental
side because the inclusive semileptonic events can serve for determining the
$B$-meson yield. Proceeding as in the previous sections, we obtain for $E_\gamma
= 1.6\,$GeV
\begin{equation}
R^{\rm SM}_\gamma \equiv 
\left({\mathcal B}^{\rm SM}_{s\gamma}+{\mathcal B}^{\rm SM}_{d\gamma}\right)/{\mathcal B}_{c\ell\nu}
= (3.31 \pm 0.22) \times 10^{-3}. 
\end{equation}
The relative uncertainties are identical to those in ${\mathcal
B}_{s\gamma}$ (as given below Eq.~(\ref{bsgsm})), except for the
parametric one which amounts to $\{+1.2,-1.7\}\%$ including the effect of
$m_b/m_q$. The gain in the overall theory uncertainty is hardly noticeable,
but this may change with the future progress in determining the perturbative
and nonperturbative corrections.

\section{Beyond-SM effects \label{sec:np}}

In most of the new-physics scenarios considered in the literature, beyond-SM
effects on ${\mathcal B}_{s\gamma}$ are driven by new additive contributions
to the Wilson coefficients of the dipole operators at the matching scale 
$\mu_0$. Denoting such contributions by $\Delta C_{7,8}$ and setting
$\mu_0$ to $160\,$GeV, we find
\begin{eqnarray} 
{\mathcal B}_{s\gamma}\times 10^4 &=& (3.36 \pm 0.23) - 8.22\,\Delta C_7 - 1.99\,\Delta C_8,\nonumber\\[2mm] 
R_\gamma\times 10^3 &=& (3.31 \pm 0.22) - 8.05\,\Delta C_7 - 1.94\,\Delta C_8.\hspace{5mm}
\end{eqnarray}
The above expressions are linearized; i.e., it is assumed that the quadratic
terms in $\Delta C_{7,8}$ are negligible when they enter with ${\mathcal O}(1)$
coefficients into the above equations.  If they are not, a detailed analysis of
QCD corrections in the considered beyond-SM scenario is necessary.

Such an analysis is available in the THDM~II~\cite{Abbott:1979dt} for which
the NLO~\cite{Ciuchini:1997xe,Borzumati:1998tg,Borzumati:1998nx} and
NNLO~\cite{Hermann:2012fc} corrections to $\Delta C_{7,8}$ are known. They
are always negative and remain practically independent of the vacuum
expectation value ratio $\tan\beta$ when $\tan\beta \gsim 2$. Sending
$\tan\beta$ to infinity in the expressions for $\Delta C_{7,8}$, we find the
following updated bounds from ${\mathcal B}_{s\gamma}$ on the charged Higgs
boson mass in this model
\begin{eqnarray}
M_{H^{\pm}} &>& 480\,\mbox{GeV~~ at 95\%~C.L.}\, ,\nonumber\\[2mm]
M_{H^{\pm}} &>& 358\,\mbox{GeV~~ at 99\%~C.L.}
\end{eqnarray}
For $\tan\beta \lsim 2$ the bounds become considerably stronger, but at the
same time other observables provide competitive
limits~\cite{Eberhardt:2013uba}. In the supersymmetric case, in which the
charged scalar and the neutral pseudoscalar tend to be almost degenerate, the
current direct search bounds~\cite{Khachatryan:2014wca,Aad:2014vgg} exceed
$500\,$GeV for $\tan\beta \gsim 20$.

\section{Summary\label{sec:sum}} 

We presented an updated prediction for ${\mathcal B}_{s\gamma}$ in the SM
taking into account all the perturbative and nonperturbative effects worked
out after the 2006 publication~\cite{Misiak:2006zs} of the first NNLO estimate
for this quantity. Our current analysis supersedes the one of
Ref.~\cite{Misiak:2006zs}.

Some of the ${\mathcal O}(\alpha_s^2)$ corrections are still interpolated in
$m_c$, but the $m_c=0$ boundary condition now comes from an explicit
calculation. Despite this improvement, the interpolation uncertainty cannot be
reduced because the interpolated correction is sizeable.  Future progress
requires extending the calculation of $G_{(1,2)7}$ to arbitrary $m_c$, which
is considered a difficult but manageable task. It would amount to evaluating
the same propagator diagrams with unitarity cuts as in
Ref.~\cite{Czakon:2015exa}, but for arbitrary $m_c$ rather than just for
$m_c=0$. Several hundreds of four-loop two-scale master integrals would need
to be calculated. For this purpose, one could numerically solve differential
equations in the variable $z = m_c^2/m_b^2$. The necessary boundary conditions
at $z \gg 1$ could be found from asymptotic expansions in this
limit. Determining such boundary conditions involves only three-loop
single-scale propagator integrals. They are likely much simpler than the
four-loop single-scale ones in Ref.~\cite{Czakon:2015exa}.

In parallel, one should investigate whether nonperturbative uncertainties can
be suppressed by combining lattice inputs with measurements of observables
like the $CP$- or isospin asymmetries in $\bar B \to X_q\gamma$. In the
analysis of Ref.~\cite{Benzke:2010js}, nonperturbative effects have been
parametrized in terms of the so-called subleading shape functions, i.e., matrix
elements of nonlocal operators between the $B$-meson states at
rest. Determining such functions directly seems to remain beyond the current
lattice capabilities. However, constraints on them can be derived from matrix
elements of local operators, the same ones that matter for the extraction of
$|V_{cb}|$ from ${\mathcal B}_{c\ell\nu}$~\cite{Alberti:2014yda}. The
higher-dimensional operator matrix elements are practically unconstrained by
the data. Any lattice estimates of them could help to suppress the
nonperturbative uncertainties in both $|V_{cb}|$ and ${\mathcal
B}_{s\gamma}$.

The main outcome of our current analysis is an upwards shift by
around $6.4\%$ in the central value of ${\mathcal B}^{\rm SM}_{s\gamma}$. It
originates mainly from fixing the $m_c=0$ boundary ($+3\%$) and including the
complete NNLO BLM corrections to the three- and four-body final state channels
($+2\%$). Both effects are within the previously~\cite{Misiak:2006zs}
estimated interpolation ($\pm 3\%$) and higher-order ($\pm 3\%$)
uncertainties, respectively.  Nevertheless, the obtained ${\mathcal
O}(1\sigma)$ increase of the central value is an important one, especially in
the context of constraining beyond-SM theories. The new four-loop calculation
of the $m_c=0$ boundary in Ref.~\cite{Czakon:2015exa} improves an essential
point in the analysis, and brings the estimated NNLO effects under much better
control.

Since ${\mathcal B}^{\rm SM}_{s\gamma}$ is now closer to ${\mathcal B}^{\rm
exp}_{s\gamma}$ (but still ${\mathcal B}^{\rm SM}_{s\gamma} < {\mathcal
B}^{\rm exp}_{s\gamma}$), the bound on $M_{H^{\pm}}$ in the THDM~II becomes
significantly stronger. The 95\%C.L. one grows by $120\,$GeV with respect
to its previous evaluation in Ref.~\cite{Hermann:2012fc} (cf. ``note added''
there).  For moderate values of $\tan\beta$, no other available measurement
constrains $M_{H^{\pm}}$ in a more efficient manner.

We supplemented our analysis with new NNLO predictions for ${\mathcal
B}_{d\gamma}$ and for the ratio
$R_\gamma = \left({\mathcal B}_{s\gamma}+{\mathcal B}_{d\gamma}\right)/{\mathcal B}_{c\ell\nu}$
where correlated uncertainties are treated in a consistent manner. The
ratio $R_\gamma$ may serve in the future as a more convenient observable for
testing beyond-SM theories with minimal flavor violation.

\begin{acknowledgments} 
We acknowledge partial support from the 
Deutsche Forschungsgemeinschaft (DFG) within the research unit FOR 1873 (QFET)                            
and within the Sonderforschungsbereich Transregio~9 ``Computational Particle Physics,''               
from the State Committee of Science of Armenia
  Program No.~13-1c153 and Volkswagen Stiftung Program No.~86426,                                    
from the Swiss National Science Foundation,                                                           
from the National Science Centre (Poland) research project, Decision No.~DEC-2014/13/B/ST2/03969,     
from the U.S.~Department of Energy, Division of High Energy Physics, under Contract DE-AC02-06CH11357,  
from the U.S.~National Science Foundation under Grant No.~PHY-1417354,                                  
and from MIUR under Contract No.~2010YJ2NYW 006.                                                      
%
%
\end{acknowledgments}

\end{document}